\documentclass[seceq]{ptptex}

\title{The $U(1)$ Topological Gauge Field Theory for Topological Defects
in Liquid Crystals}

\author{Yi-shi {\sc Duan}$^{1,}$  Li  {\sc Zhao} $^{1,}$ \footnote{E-mail:
zhl03@st.lzu.edu.cn} Xin-hui {\sc Zhang}$^{1,}$ and Tie-yan {\sc
Si} $^2$ }

\inst{$^1$Institute of Theoretical Physics, Lanzhou University,
Lanzhou, 730000, P. R. China
\\
$^2$Institute of Theoretical Physics, Chinese Academy of Science,
Beijing, 100080, P. R. China }

\abst{A novel $U(1)$ topological gauge field theory for
topological defects in liquid crystals is constructed by
considering the $U(1)$ gauge field is invariant under the director
inversion. Via the $U(1)$ gauge potential decomposition theory and
the $\phi$-mapping topological current theory, the decomposition
expression of $U(1)$ gauge field  and the unified topological
current for monopoles and strings in liquid crystals are obtained.
It is revealed that monopoles and strings are located in different
spatial dimensions and their topological charges are just the
winding numbers of $\phi$-mapping. }

\begin{document}

\maketitle

\section{Introduction}

Topological defects have attracted a lot of interest in liquid
crystals \cite{Toth,Andrienko, Kleman} and play an important role
in the static structures and dynamical behavior of liquid
crystals. \cite{Grollau,Denniston} Because of the soft-liquid
character and optic transparency of liquid crystals, a rich
variety of topological defects are exhibited, such as
dislocations, disclinations, point defects, global defects.
Theoretically, these topological defects have been classified by
the homotopy group, which provides a natural language for the
classification of defects in ordered media. Work in this field was
originated by Finkelstein, \cite{Finkelstein1966} Toulouse and
Kl\'{e}man \cite{Toulouse1976}$et\; al$ and  was summarized by
Mermin, \cite{Mermin1979}Anderson \cite{Anderson1984} and Bray
.\cite{Bray1994} Consider the symmetry breakdown of Lie group $G
\rightarrow H$, taking for $G$ the simply connected cover of the
initial Lie symmetry and $H \subset G$ its unbroken subgroup.
Monopoles lie on the vacuum manifold with nontrivial
$\pi_{2}(G/H)$. Strings occur in theories with nontrivial
$\pi_{1}(G/H)$, domain walls occur with nontrivial $\pi_{0}(G/H)$
and textures in theories with nontrivial $\pi_{3}(G/H)$.  In the
case of biaxial liquid crystals in three-dimensional space, the
vacuum manifold is a sphere, where diametrically opposite points
have to be identified because of the equivalence of director
$\mathbf{d}$ and $-\mathbf{d}$. \cite{Gennes1993} This is the
projection plane $P_{2}$, whose group element is $G/H\cong
SO(3)/D_{2}$. \cite{Michel1977} In addition to their importance in
ordered media, these classifications are also relevant to problems
in cosmological structure formation. However, this earlier work is
mainly concerned with the topological classification
of individual defects, and the topological gauge field theory for liquid crystals is not established yet.\\
\indent In the present paper, we will construct a  $U(1)$
topological gauge field theory for topological defects in liquid
crystals, in which the $U(1)$ gauge field is invariant under the
director inversion $\mathbf{d}\rightarrow -\mathbf{d}$.  We will
derive a unified topological current for monopoles and strings in
liquid crystals from the $U(1)$ gauge field tensor and show that
monopoles and strings are generated from the singularities of the
director field in different spatial dimensions. This work is based
on the so-called gauge potential decomposition theory and
$\phi$-mapping topological current theory.
\cite{Meng1993}\cite{Duan1984} \\
\indent This paper is arranged as follows. In Sec. \ref{section1},
we investigate the decomposition expression of the $U(1)$ gauge
potential in liquid crystals. In Sec. \ref{section2}, the
topological current for monopoles and strings in liquid crystals
is presented. In Sec. \ref{section3}, we show that the topological
charges of monopoles and strings are just the winging numbers. In
the final section, we provide a brief conclusion and outlook for
the further work.
\section{The $U(1)$ gauge potential decomposition in terms of the director
           }\label{section1}

As is well known, the director $\mathbf{d}$ has unit length by
definition and is to be identified with $-\mathbf{d}$ in liquid
crystals. In this sense the director becomes a unit double-valued
vector \cite{Blaha1976} and generalizes the former concept of a
single-valued vector. Generally, the topological characteristics
of a manifold are represented by the properties of a smooth vector
field on it, or, in other words, the smooth vector fields carry
the topological information of a manifold. In the single-valued
vector field system, the gauge potential decomposition theory is
established by means of the unit vector field $\mathbf{n}$.
\cite{Zhang1998} In liquid crystals, its inner structure should be
based on the director field $\mathbf{d}$. In the case of the
2-dimensional director, the director inversion corresponds to the
$U(1)$ gauge transformation, which inspires us to study the $U(1)$
gauge potential decomposition in terms of the unit director field
on the vacuum manifold.\\
\indent We first study the $U(1)$ gauge potential decomposition in
the single-valued vector field system, which reveals the inner
structure of the $U(1)$ gauge potential in terms of $\mathbf{n}$.
It is known that the $U(1)$ gauge covariant derivative and its
complex conjugate are given by
\begin{eqnarray}
D_{\mu}\psi&=&\partial_{\mu}\psi-iA_{\mu}\psi , \label{covariant
derivative}
\\(D_{\mu}\psi)^{*}&=&\partial_{\mu}\psi^{*}+iA_{\mu}\psi^{*},
\label{complex conjugate}
\end{eqnarray}
where $A_{\mu}$ is the $U(1)$ gauge potential (connection) and
$\psi$ is the complex wave function in the single-valued vector
field. And the $U(1)$ gauge field tensor (curvature) is defined by
\begin{equation}
F_{\mu\nu}=\partial_{\mu}
A_{\nu}-\partial_{\nu}A_{\mu},\label{gauge field tensor1}
\end{equation}
which is invariant under the $U(1)$ gauge transformation:
$A_{\mu}^{'}(x)=A_{\mu}(x)+\partial_{\mu}\theta(x),$ where
$\theta(x)\in R$ is a phase factor denoting the $U(1)$ gauge
transformation. Multiplying $\psi^{*}$ with Eq. (\ref{covariant
derivative}) and Eq. (\ref{complex conjugate}) with $\psi$, it is
easy to find the decomposition expression of the $U(1)$ gauge
potential
\begin{equation}
A_{\mu}=\frac{1}{2i\psi^{*}\psi}(\psi^{*}\partial_{\mu}\psi-\partial_{\mu}\psi^{*}\psi)
-\frac{1}{2i\psi^{*}\psi}(\psi^{*}D_{\mu}\psi-(D_{\mu}\psi)^{*}\psi).\label{gauge
potential}
\end{equation}
To study the inner structure of the above expression. The complex
wave function $\psi(x)$ can be regarded as a complex
representation of  a $2$-dimensional vector field
$\vec{\psi}=(\varphi^{1},\varphi^{2})$ over the vacuum manifold,
i.e.,
\begin{equation}
\psi(x)=\varphi^{1}(x)+i\varphi^{2}(x),
\end{equation}
where $\varphi^{1}(\vec{x})$ and $\varphi^{2}(\vec{x})$ are two
real functions. The $2$-dimensional unit vector $\mathbf{n}$ is
naturally expressed as
\begin{equation}
n^{a}=\frac{\varphi^{a}}{\|\varphi\|},\;(a=1,2),
\end{equation}
where $\|\varphi\|^{2}=\varphi^{a}\varphi^{a}=\psi^{*}\psi$. Then
Eq. (\ref{gauge potential}) can be simplified as
 \begin{equation}
A_{\mu}=\epsilon^{ab}n^{a}\partial_{\mu}n^{b}-\epsilon^{ab}n^{a}D_{\mu}n^{b}.\label{gauge
potential'}
\end{equation}
Let $k^a=\epsilon ^{ab}n^b $ be another $2$-dimensional unit
vector which is orthogonal to $n^a$:
 $k^an^a=0,\;k^ak^a=1, $ then using $n^a$ and $k^a$, Eq. (\ref{gauge
potential'}) is rewritten as
\begin{equation}
A_{\mu}=k^a\partial _{\mu}n^a-k^aD_{\mu}n^a. \label{2-complvecdec}
\end{equation}
Suppose $u^a$ be another unit vector field satisfying:
$D_iu^a=0\;\;(u^au^a=1)$ and be expressed as $u^a=\cos \theta
n^a+\sin \theta k^a$, it can be proved that $-k^aD_in^a=\partial
_i\theta.$ We see that the second term of (\ref{gauge
potential'}), $\partial _i\theta $, behaves as a $U(1)$ gauge
transformation of $A_{\mu}$, which contributes nothing to the
gauge field tensor $F_{\mu\nu}$ defined by Eq. (\ref {gauge  field
tensor1}) and can be ignored in $U(1)$ decomposition theory.
\cite{Zhang1998}Therefore the decomposition of $U(1)$ gauge
potential is simplified as
\begin{equation}
A_{\mu}=\frac{1}{2i}\frac 1{\psi ^{*}\psi }(\psi ^{*}\partial
_i\psi -\partial _i\psi ^{*}\psi
)=\epsilon^{ab}n^{a}\partial_{\mu}n^{b}. \label{2-decomU1}
\end{equation}
which means that the $U(1)$ gauge potential $A_{\mu}$ possesses an
inner structure in terms of the $2$-dimensional unit vector
$\mathbf{n}$. In superconductivity theory, the form of
(\ref{2-decomU1}) actually corresponds to the London relation
\cite{Feynman} and  is a fundamental expression in $%
U(1)$ topological quantum mechanics. Thus, the $U(1)$ gauge field
tensor $F_{\mu\nu}$ in Eq. (\ref{gauge  field tensor1}) is changed
into
\begin{equation}
F_{\mu\nu}=2\epsilon^{ab}\partial_{\mu}n^{a}\partial_{\nu}
n^{b}.\label{gauge field tensor}
\end{equation}
From Eqs. (\ref{2-decomU1}) and (\ref{gauge field tensor}), it is
easy to see that the $U(1)$ gauge field is invariant when the unit
vector field $\mathbf{n}$ transforms into $-\mathbf{n}$.\\
\indent Now we generalize the above $U(1)$ gauge potential
decomposition to the case of the double-valued vector
system-----liquid crystals. We will show that the $U(1)$ gauge
field in liquid crystals possesses an inner structure in terms of
the $2$-dimensional director $\mathbf{d}$. Note that, in nematic
liquid crystals, the director inversion
$\mathbf{d}\rightarrow-\mathbf{d}$ corresponds to the $U(1)$ gauge
transformation,
\begin{equation}
{d}^{'}=e^{i\alpha(\vec{x})}{d} ,\quad
A_{\mu}^{'}=A_{\mu}+\partial_{\mu}\alpha(\vec{x})\label{gauge
transformation}
\end{equation}
with the phase factor $\alpha=\pi$. Then the $U(1)$ gauge
potential $A_{\mu}^{'}$ equals to $A_{\mu}$ under
$\mathbf{d}\rightarrow-\mathbf{d}$, which means the director
inversion leaves the gauge field unchanged and does not influence
the physics of the system. \cite{Gennes1993} Because of the
invariance of the above decomposition expressions in Eqs.
(\ref{2-decomU1}) and (\ref{gauge field tensor}) under
$\mathbf{n\rightarrow -\mathbf{n}}$, the single-valued vector
$\mathbf{n}$  can be substituted by the double-valued director
$\mathbf{d}$, i.e.,
\begin{equation}
A_{\mu}=\epsilon^{ab}d^{a}\partial_{\mu}d^{b}, \quad
F_{\mu\nu}=2\epsilon^{ab}\partial_{\mu}d^{a}\partial_{\nu} d^{b},
\label{gauge field tensor'}
\end{equation}
which are naturally invariant under the director inversion
$\mathbf{d}\rightarrow -\mathbf{d}$. As a result, in liquid
crystals the $U(1)$ gauge potential $A_{\mu}$ and gauge field
tensor $F_{\mu\nu}$ can be decomposed in terms of the
$2$-dimensional director ${d^{a}}\;(a=1,2)$.\\
\indent In the following section, using the $\phi$-mapping
topological current theory, we show that there is a topological
current which can be derived from $F_{\mu\nu}$, and topological
defects structure is just inhering in this topological current.

\section{The topological structure of monopoles and strings in liquid crystals
}\label{section2}

Although a rich array of topological defects may appear in liquid
crystals, the exotic types of defects, i.e., domain walls and
textures, are more energetic and therefore rare in appearance.
Here we focus on the monopole and string configurations with the
defects' dimensionality $0$ and $1$, respectively. The study of
the topological defects is often involved in a $n$-dimensional
order parameter associated with the $d$-dimensional space. For the
$2$-dimensional director case, the defects are point defects
(monopoles) for the spatial dimensionality $d=2$, line defects
(strings) for $d=3$. More generally, for $n=d$, one has point
defects ; for $n=d-1$, one generates line defects.

Let us consider the $\phi$-mapping
 $\phi^{a}:x\rightarrow\phi^{a}(x)\;(a=1,2)$ with $x=(x^{1},x^{2},\cdots,x^{i})$
the local coordinate in real physical space,
 which is called the order parameter of topological defects in liquid
 crystals. Since the unit director
 $\mathbf{d}$ is  parallel to the order parameter $\vec{\phi}=(\phi^{1},\phi^{2})$, it
 can be further expressed as direction field of order parameter
at the places $\phi^{a}(x)\neq 0$, i.e.,
\begin{equation} d^{a}=\frac{\phi^{a}}{\|\phi\|},\quad \|\phi\|^{2}=\phi^{a}\phi^{a},\quad d^{a}d^{a}=1,
\quad a=1,2.\label{unit vector field}
\end{equation}
From Eq. (\ref{unit vector field}), one can see that the director
inversion $\mathbf{d}\rightarrow -\mathbf{d}$ corresponds to the
transformation of ${\vec{\phi}}\rightarrow -{\vec{\phi}}$
 and the zeros of $\vec{\phi}$ are just the singularities of
$\mathbf{d}$ at which the director field is indefinite. In liquid
crystals, it is known that topological defects locate at the
singularities of the director field $\mathbf{d}$. This property
will be reflected mathematically in the following.

From the expression of the gauge field tensor $F_{\mu\nu}$ in Eq.
(\ref{gauge field tensor'}), the topological current for monopoles
and strings in  liquid crystals is introduced by
\begin{equation}
j^{i}=\frac{1}{4\pi}\epsilon^{ijk}F_{jk}=\frac{1}{2\pi}
\epsilon^{ijk}\epsilon_{ab}\partial_{j}d^{a}\partial_{k}d^{b},\quad
i,j,k=1,2,3.\label{topological current}
\end{equation}
where for monopoles $i,j,k$ denote the $(2+1)$-dimensional
space-time coordinates with $x^{1}$ the time parameter, and for
strings, $i,j,k$ denote the $3$-dimensional space coordinates. The
topological current in the above equation is the special case of
the $\phi$-mapping topological current theory. \cite{Duan1984}
 Then using
$\partial_{i}d^{a}=\partial_{i}({{\phi}^{a}}/{\|{\phi}\|})
={\partial_{i}{\phi}^{a}}/{\|{\phi}\|}+{\phi}^{a}\partial_{i}({1}/{\|{\phi}\|})$
and the Green function equation in ${{\phi}}$ space,
$\partial_{a}\partial_{a}ln\|{\phi}\|=2\pi\delta^{2}({\vec{\phi}})
(\partial_{a}=\frac{\partial}{\partial{\phi}^{a}})$, we obtain the
$\delta$-function like topological current \cite{Duan1998}
\begin{equation}
j^{i}=\delta^{2}({\vec{\phi}})D^{i}(\frac{{\phi}}{x}),\label{a8}
\end{equation}
where
$D^{i}(\frac{{\phi}}{x})=\frac{1}{2}\epsilon^{ijk}\epsilon^{ab}\partial_{j}
{\phi}^{a}\partial_{k}{\phi}^{b}$ is the Jacobian vector. Since
the $\delta$-function and the Jacobian vector
$D^{i}(\frac{{\phi}}{x})$ satisfy
\begin{eqnarray}\delta(\vec{\phi})=\delta(-\vec{\phi})\quad and
\quad D^{i}(\frac{{\phi}}{x})=D^{i}(-\frac{{\phi}}{x}),
\end{eqnarray} the topological current in Eq. (\ref{a8})
 is invariant under the order parameter transformation
${\vec{\phi}}\rightarrow-{\vec{\phi}}$, which means the
topological current is also applicable for liquid crystals.

The expression of Eq. (\ref{a8}) provides an important conclusion
\begin{equation}
j^i\left\{
\begin{array}{l}
=0,\quad if \;and\;only\;if\;{\vec{\phi}}\neq 0, \\
\neq 0,\quad if \;and\;only\;if\;{\vec{\phi}}=0,
\end{array}
\right.
\end{equation}
so it is necessary to study the zero points of ${\vec{\phi}}$ to
determine the non-zero solutions of $j^i.$ Suppose that the order
parameter $\phi^{a}(x)$ processes $N$ isolated zeros, according to
the implicit function theory, \cite {Goursat1904}  under the
regular condition
\begin{equation}
D^{i}({\phi} /x)\neq 0,  \label{b11}
\end{equation}
the general solutions of
\begin{equation}
\left\{
\begin{array}{l}
{\phi}^1(x^1,x^2,
x^{3})=0, \\
{\phi}^2(x^1,x^2,x^{3})=0,
\end{array}
\right.
\end{equation}
can be point-like form: $z_{k}=z_k(t)$ $(k=1,\;2,\;\cdots,\;N)$
located in the $(2+1)$-dimensional space-time or string-like form
$L_{k}$: $z_{k}=z_k(s)\;$ (s is a line parameter) located in the
$3$-dimensional space. These singular solutions are just monopoles
or strings generated from the zeroes of the order parameter field
$\vec{\phi}$, i.e., the singularities of the director field
$\mathbf{d}$.\\
\indent We notice that $\delta(\vec{\phi})=\infty$ for
$x=z_{k}(t)$ or $z_{k}(s)$, and vanishes outside these zeros.
According to the $\delta $-function theory, \cite{Schouten1951}
$\delta(\vec{\phi})$ can be expanded as
\begin{equation}
\delta(\vec{\phi})=\left\{
\begin{array}{l}
\sum_{k=1}^{N}\frac{\beta _k}{{\left|
D(\frac{{\phi}}{x})\right| _{z _k}}} \delta(x-z_{k}(t))\quad for \;monopoles,\\
\sum_{k=1}^{N} \int_{L_{k}}\frac{\beta _k}{{\left|
D(\frac{{\phi}}{u})\right| _{\Sigma _k}}}\delta(x-z_{k}(s))ds\quad
for \;strings, \label{delta function theory}
\end{array}
\right.
\end{equation}
where
$D(\frac{\phi}{x})|_{z_{k}}=\frac{1}{2}\epsilon^{jk}\epsilon_{mn}\frac{\partial\phi^{m}}{\partial
x^{j}} \frac{\partial\phi^{n}}{\partial x^{k}}$;
$D(\frac{\phi}{u})|_{\Sigma_{k}}=\frac{1}{2}\epsilon^{jk}\epsilon_{mn}\frac{\partial\phi^{m}}{\partial
u^{j}} \frac{\partial\phi^{n}}{\partial u^{k}}$ and $\Sigma_{k}$
is the $k$-th planer element transversal to the string $L_{k}$
with local coordinates $(u^{1},u^{2})$. The positive $\beta_{k}$
is the Hopf index of $\phi$-mapping. In liquid crystals, due to
the equivalence of the director $\mathbf{d}$ and $\mathbf{-d}$,
the Hopf index can be integers and half-integers. Meanwhile the
general velocity $V^{i}$ of the $k$th monopole and string  is
respectively given by, \cite{Duan1998}
\begin{equation}
\left. \frac{dz^i_{k}}{dt}\right| _{z_k(t)}=\left. \frac{D^i({\phi}/x)}{D({\phi} /x)}%
\right| _{z_k}, \quad \left. \frac{dz^i_{k}}{ds}\right| _{z_k(s)}=\left. \frac{D^i({\phi}/x)}{D({\phi} /u)}%
\right| _{z_k}, \label{dx/ds}
\end{equation}
which determines the motion of the $k$th topological defect
$z_{k}$. Then applying Eqs. (\ref{a8}), (\ref{delta function
theory}) and (\ref{dx/ds}), the topological current becomes
\begin{equation}
j^{i}=\left\{
\begin{array}{l}
\sum_{k=1}^{N} \beta_{k}\eta_{k}\frac{dz_{k}^{i}}{dt}\delta(x-z_{k}(t))\quad \quad \quad for \;monopoles,\\
\sum_{k=1}^{N}
\beta_{k}\eta_{k}\int_{L_{k}}\frac{dz_{k}^{i}}{ds}\delta(x-z_{k}(s))ds\quad\quad
for \;strings, \label{topological current'}
\end{array}
\right.
\end{equation}
where $\eta _k=sgnD(\phi /x)_{z_k}=\pm 1$ (for monopoles) or $\eta
_k=sgnD(\phi /u)_{\Sigma_k}=\pm 1$ (for strings) is the Brouwer
degree of $\phi$-mapping: $\eta_{k}=+1$ corresponds to the defects
solutions, while $\eta_{k}=-1$ corresponds to anti-defects
solutions. Between the topological current and topological current
density, there is an important relation: $j^{i}=\rho V^{i}$ with
the general velocity $V^{i}$. From Eq. (\ref{topological
current'}), the topological current density $\rho$ becomes

\begin{equation}
\rho=\left\{
\begin{array}{l}
\sum_{k=1}^{N} \beta_{k}\eta_{k}\delta(x-z_{k}(t))\quad \quad for \;monopoles,\\
\sum_{k=1}^{N}
\beta_{k}\eta_{k}\int_{L_{k}}\delta(x-z_{k}(s))ds\quad\quad for
\;strings,\label{topological current density}
\end{array}
\right.
\end{equation}
This quantized current in Eq. (\ref{topological current'}) and
current density in Eq. (\ref{topological current density}) are the
unified form of the defect densities done by Liu and Mazenko.
\cite{Liu1992} In their work the topological index
$\beta_{k}\eta_{k}$ is absent and it requires additional homotopic
analysis to classify defects.

From Eq. (\ref{topological current density}), one can obtain the
topological charge of the $k$th monopole and string
\begin{equation}
Q_k=\int_{\Sigma _k}\rho d\sigma =\beta_{k}\eta_{k},
\label{topological charge}
\end{equation}
where the  surface  $\Sigma_{k}$ is a neighborhood  of $z_{k}$
with $z_{k}\neq \partial\Sigma_{k}$, $\Sigma_{k}\bigcap
\Sigma_{l}=\O$. For monopoles, the surface element $d\sigma$
satisfies: $\epsilon^{ij}d\sigma=dx^{i}\wedge dx^{j}\; (i,j=1,2)$;
while for strings, $\epsilon^{ij}d\sigma=du^{i}\wedge
du^{j}\;(i,j=1,2)$. Then the total topological charge carried by
$N$ topological defects on $\Sigma=\cup \Sigma_{k}$ is
\begin{equation}
Q=\sum_{k=1}^{N}Q_k =\sum_{k=1}^{N}\beta_{k}\eta_{k}.
\end{equation}
It is obvious to see that the topological current in Eq.
(\ref{topological current'}) represents $N$ topological defects of
which the $k$th topological defect is charged with the topological
charge $\beta_{k}\eta_{k}$. Therefore, this current describes how
the defects move and how the topological charges are distributed
for topological defects in a topologically quantized way.

Recent experiments with nematic liquid crystals have revealed
interesting knotlike vortex lines,\cite{Bowick1994} which include
the ring-like and knot-like configurations. For the topology of
the closed lines $L_{ring}$ and ${L_{knot}}$, we obtain the
topological current

\begin{equation}
j^{i}=\sum_{i=1}^{N}\beta_{r_{i}}\eta_{r_{i}}\oint_{L_{ring}}
\frac{D^{i}(\phi/x)}{D(\frac{\phi}{u})}\delta(x-z_{r_{i}}(s))ds+\sum_{i=1}^{N}
\beta_{k_{i}}\eta_{k_{i}}\oint_{L_{knot}}
\frac{D^{i}(\phi/x)}{D(\frac{\phi}{u})}\delta(x-z_{k_{i}}(s))ds,\label{knotlike
vortex lines}
\end{equation}
where  $\beta_{r_{i}}\eta_{r_{i}}$ and $\beta_{k_{i}}\eta_{k_{i}}$
are the $i$th topological number of $\vec{\phi}$ surrounding the
ring and knot like lines $L_{ring}$ and $L_{knot}$, respectively.
\section{ The winding numbers of monopoles and strings in liquid
crystals}\label{section3} Defects are classified according to
their topological  strength $S$, \cite{Semenov1999} and a more
general definition of the topological strength is the winding
number $W$ of topological defects. In liquid crystals,
half-integer values of winding numbers are also allowed ($W=\pm
\frac{1}{2},\pm1 \pm\frac{3}{2},\ldots$) because of  the ``head to
tail" invariance of the director. Defects of opposite winding
number may annihilate each other.

In the single-valued vector field system, the  winding number
$W_{k} $ is defined by the Gauss map $n$.  As to monopoles, we
definitely consider the $2$-dimensional space. Moreover, as to
strings in the system, for simplicity and without loss of
generality, we also suppose  that the director field $\mathbf{d}$
lies locally in the plane which is vertical to strings with the
local coordinates $(u^{1},u^{2})$. The Gauss map is $n:\;\partial
\Sigma_{k} \rightarrow S^2$ and the winding number $W_{k} $ is
defined by
\begin{equation}
W_{k}=\frac 1{2\pi }\int_{\partial \Sigma_{k} }n^{*}(\epsilon
_{abc}n^a\wedge dn^b ),
\end{equation}
where $n^{*}$ is the pull back of the map $n$ and  $\partial
\Sigma_{k} $ is the boundary of  $\Sigma_{k}$. The winding number
is a topological invariant and is called the degree of Gauss map
\cite{Milnor1966}. Using the Stokes' theorem, the above formula
can be written as
\begin{equation}
W_{k}=\frac{1}{2\pi}\int_{\Sigma_{k}}\epsilon_{ab}dn^{a}\wedge
dn^{b}=\frac{1}{2\pi}\int_{\Sigma _{k}}\epsilon^{ij}\epsilon_{ab}
\partial_{i}n^{a}\partial_{j}n^{b} d\sigma.
 \label{winding number}
\end{equation}
From the above equation, the winding number $W_{k}$ is an integral
invariant under the transformation $\mathbf{n}\rightarrow
\mathbf{-n}$. Therefore, the winding number $W_{k}$ for
topological defects in liquid crystals can be expressed as
\begin{equation}
W_{k}=\frac{1}{2\pi}\int_{\Sigma_{k}}\epsilon_{ab}d(d^{a})\wedge
d(d^{b})=\frac{1}{2\pi}\int_{\Sigma
_{k}}\epsilon^{ij}\epsilon_{ab}
\partial_{i}d^{a}\partial_{j}d^{b} d\sigma.\label{winding number}
\end{equation}
\\ \indent In topology, the winding number $W_{k}$ means that when the
topological defect $z_{k}$ covers $\Sigma_{k}$ in the real space
once, the unit vector  will cover the unit sphere $S^{2}$ for
$W_{k}$ times.  In the single-valued vector field system, to
return the same, the vector field $\mathbf{n}$ rotates by  $2k\pi$
around the topological defects, where $k$ is an integer. While in
liquid crystals, due to the equivalence of $\mathbf{d}$ and
$\mathbf{-{d}}$, the director field $\mathbf{d}$ only needs to
rotate by $k\pi$. By the topological meaning of the winding
number, $W_{k}$ in liquid crystals is half of that in the
single-valued vector field system, that is to say, $W_{k}$ can be
integers and half-integers, which is a generalization of our
previous integer winding number.

On the other hand, $W_{k}$ is also the topological charge $Q_{k}$
of the $k$th  topological defect. Substituting Eq.
(\ref{topological current density}) into Eq. (\ref{winding
number}), one can find the relation between the winding number
$W_{k}$, the Hopf index $\beta_{k}$ and the Brouwer degree
$\eta_{k}$
\begin{equation}
W_{k}=\int_{\Sigma_{k}}\rho d\sigma =\beta_{k}\eta_{k},
\end{equation}
which shows that the winding number $W_{k}$  is just the
topological charge $Q_{k}$ in Eq. (\ref{topological charge}) of
the $k$th topological defect. For liquid crystals with a set of
topological defects, the total winding number is
\begin{equation}
W=\sum_{k=1}W_{k}=
\sum_{k=1}\beta_{k}\eta_{k}=W_{+}-W_{-},\label{topological
quantization}
\end{equation}
where $W_{+}$ and $W_{-}$ are the total winding number of defects
and anti-defects. The above expression in Eq. (\ref{topological
quantization}) naturally arrives at the conclusion that, in liquid
crystals, each isolated topological defect created from the
singularities  of the director is characterized by the topological
number $W_{k} $.  As some corollaries of Eq. (\ref{topological
quantization}), the following two points are obtained. (i) For a
certain surface, the total winding number of topological defects
is not arbitrary but a topological invariant-----the Euler
characteristic on the vacuum manifold obtained from the
Gauss-Bonnet-Chern theorem and Hopf index theorem. (ii) In liquid
crystals, since the free energy is proportional to the square of
winding number, there exists in general the monopoles of winding
number $\pm 1$ and strings of winding number $\pm\frac{ 1}{2}$
with the principle of least free energy.

\section{Conclusion}\label{section4}
In summary, in the light of  the gauge potential decomposition
theory and the $\phi$-mapping topological current theory, we
construct a $U(1)$ topological gauge field theory for monopoles
and strings in liquid crystals, in which the $U(1)$ gauge field is
invariant under the director inversion. It is revealed that in
liquid crystals the $U(1)$ gauge potential and $U(1)$ gauge field
tensor can be decomposed in terms of the $2$-dimensional director
$\mathbf{d}$, which is essential to study the topological
properties of monopoles and strings located at the singularities
of the director field. We derive a unified topological current for
monopoles and strings from the $U(1)$ gauge field tensor and show
that the topological current takes the form of $\delta$-function.
Furthermore, the topological charge carried by monopole and string
is just the winding number labelled by the Hopf index and Brouwer
degree.  The theory formulated in this paper is a new concept for
topological defects in liquid crystals.

At last there are two points which should be stressed. Firstly,
 when the regular condition Eq. (\ref{b11}) fails caused by
 the external field (e.g. the magnetic field, the electric field),
 the bifurcation processes of topological defects will occur.
 Secondly, the topological current of the knotlike vortex lines
  in liquid crystals is given in Eq. (\ref{knotlike vortex lines}),
  which will give us a insight into the topological essence of the knotlike vortex lines and
be detailed in our further work.

\section*{Acknowledgements}
This work was supported by the National Natural Science Foundation
and Doctor Education Fund of the Educational Department of the
People's Republic of China.

%

\end{document}